# NeRTCAM: CAM-Based CMOS Implementation of Reference Frames for Neuromorphic Processors

Harideep Nair, William Leyman, Agastya Sampath, Quinn Jacobson, and John Paul Shen
*ECE Department, Carnegie Mellon University*
{hpnair, wleyman, agastyas, qjacobso, jpshen}@andrew.cmu.edu

*Abstract*—Neuromorphic architectures mimicking biological neural networks have been proposed as a much more efficient alternative to conventional von Neumann architectures for the exploding compute demands of AI workloads. Recent neuroscience theory on intelligence suggests that Cortical Columns (CCs) are the fundamental compute units in the neocortex and intelligence arises from CC's ability to store, predict and infer information via structured Reference Frames (RFs). Based on this theory, recent works have demonstrated brain-like visual object recognition using software simulation. Our work is the first attempt towards direct CMOS implementation of Reference Frames for building CC-based neuromorphic processors. We propose *NeRTCAM* (Neuromorphic Reverse Ternary Content Addressable Memory), a CAM-based building block that supports the key operations (store, predict, infer) required to perform inference using RFs. NeRTCAM architecture is presented in detail including its key components. All designs are implemented in SystemVerilog and synthesized in 7nm CMOS, and hardware complexity scaling is evaluated for varying storage sizes. NeRTCAM system for biologically motivated MNIST inference with a storage size of 1024 entries incurs just 0.15 mm$^2$ area, 400 mW power and 9.18 $\mu$s critical path latency, demonstrating the feasibility of direct CMOS implementation of CAM-based Reference Frames.

*Index Terms*—Neuromorphic Processors, Cortical Columns Computing, Reference Frame, Content Addressable Memory

## I. INTRODUCTION

Over the last decade, rapid advancements in Artificial Intelligence (AI), particularly Deep Learning (DL), have revolutionized various domains, ranging from computer vision and natural language processing to robotics and autonomous systems. Current implementations of AI algorithms, particularly deep neural networks (DNNs), typically involve processing massive amounts of data, employing high-dimensional tensor processing and gradient-based backpropagation. Traditional computing systems based on Turing computation model and von Neumann architecture driven by Moore's Law have served as the primary hardware substrate fueling this AI juggernaut. However, the exponentially increasing computational demand, and thereby power consumption, pose significant challenges for continued sustenance of this trend [1]–[3].

Neuromorphic computing architecture addresses these challenges by mimicking the fundamental principles of biological neural networks in the human brain. One such approach, called *Cortical Columns Computing Systems*, was recently introduced in [4], which suggests neuromorphic architectures can be developed based on the fundamental computational units within the brain, Cortical Columns (CCs). Such computing systems can potentially achieve brain-like energy efficiency and brain-like capabilities, including online continual learning.

In his recent book titled *"A Thousand Brains: A New Theory of Intelligence"* [5], Jeff Hawkins proposes a new theory on intelligence suggesting that cortical columns in the neocortex model sensory information and knowledge in structured *Reference Frames* (RFs) and continuously predict and update the stored models as the sensing agent moves in an environment. Hawkins postulates that such a Reference Frame, inspired from biological grid cells in the entorhinal cortex [6] and place cells in the hippocampus [7], implements the common universal algorithm for achieving intelligence within each cortical column, irrespective of the sensory modality feeding that particular cortical column.

Following this, Hawkins' colleagues from Numenta successfully demonstrated visual object recognition on MNIST using a grid cell-based network, called GridCellNet [8], that learns a digit by associating features at certain locations within an implicit RF and classifies digits by obtaining sequential input features through movement in this RF. Prior works [9], [10] have illustrated the important roles of RFs in cortical columns, and the potential applications of this new theory based on software simulation results. However, currently no prior work has investigated the feasibility and the potential of direct implementation of Reference Frames in conventional off-the-shelf digital CMOS technology.

Based on Hawkins' theory, authors in [4] model a Cortical Column as consisting of an Agent and a Reference Frame, wherein the RF maintains a map of the sensory information collected by the agent and the agent then achieves goal-oriented behaviors using input sensory features along with information from the RF. In this model, intelligence arises as a result of "movement" in the brain, wherein the system learns to associate "features" (semi-unique identifiers for tangible structures on an object, or abstract concepts forming a whole idea in a cognitive map) with "locations" in an RF. This model of intelligence necessitates having a form of associative lookup for features given locations, and vice versa. Hence, in this work, we propose a content addressable memory (CAM) based microarchitecture as a step towards direct CMOS implementation of the RF. Specifically, we propose the Neuromorphic Reverse Ternary Content Addressable Memory (NeRTCAM) targeting inference using RFs. Key contributions include:

- We propose the Neuromorphic Reverse Ternary CAM (NeRTCAM) - a CAM-based system that supports the key

operations needed to perform inference using a Reference Frame (RF). To the best of our knowledge, this is the first work that illustrates the hardware implementation feasibility of RF for cortical column computing.

- We specify the system architecture for NeRTCAM, including input commands (macro-operations or macro-ops) as well as internal micro-operations (or micro-ops).
- We build on the specialized CAM called ternary CAM (TCAM). Instead of matching binary query with stored ternary entries consisting of 'don't-care' bits as in TCAM, we match queries consisting of 'don't-care' bits with binary stored entries, hence named *Reverse TCAM*.
- We use MNIST as an example benchmark from [8] and generate 7nm post-synthesis power-performance-area (PPA) results for NeRTCAM. We analyze implementation results with varying storage sizes and distribution of hardware complexity across the system components.

This paper is organized as follows. Two key bodies of prior work are discussed in Section II, followed by description of NeRTCAM system architecture in Section III. Section IV provides specific implementation details for each of the NeRTCAM system components. Post-synthesis PPA evaluation and results are presented in Section V followed by summary of our key conclusions and future work in Section VI.

## II. RELATED WORKS

### A. Cortical Columns Computing Systems

Neuromorphic spiking neural networks (SNNs) offer an alternative approach to AI compute, based on brain-inspired principles and brain-like architecture. Smith's Temporal Neural Networks (TNNs) [11] emerged recently as a more biologically plausible approach utilizing efficient temporal encoding, local Spike Timing Dependent Plasticity (STDP) learning and neocortical hierarchy comprising of dendrites, synapses, neurons, (mini)columns, and layers. Prior works have also developed a microarchitecture model and custom building block macros for implementing highly efficient TNN designs [12], [13]. Authors in these works illustrated that single TNN mini-columns can perform unsupervised time-series clustering [14], [15] within 40 $\mu$W power while outperforming majority of the state-of-the-art alternatives. Further, multi-layer TNNs can achieve state-of-the-art 99% accuracy on MNIST digit recognition within 18 mW power, while enabling fast on-chip and online continuous learning.

Further, Hawkins' new theory on intelligence, The Thousand Brains Theory [5], proposes *Cortical Columns (CCs)* as the fundamental neocortical processing units. Human brain has around 150,000 such CCs. Higher forms of intelligence result from larger number of CCs. The neocortex gains its intelligence through CC's ability to model sensory information in structured Reference Frames (RFs), and learn models/maps of objects through predict-sense-update feedback loops. Reference Frames are directly inspired from the biological grid cells in the entorhinal cortex and place cells in the hippocampus, that support spatial navigation for humans in the real world.

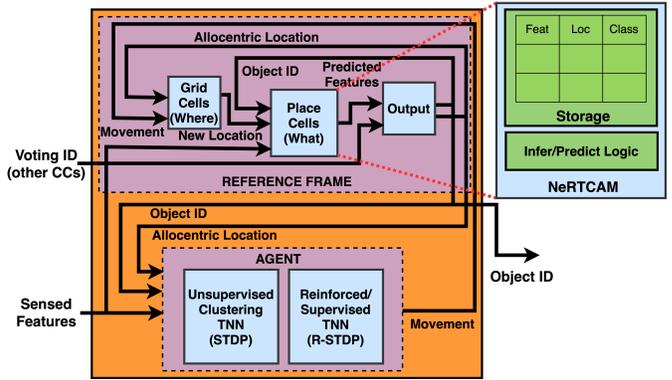

Fig. 1. NeRTCAM as part of Reference Frame (RF) within a Cortical Column (*adapted from [4]*). It is used as a building block for RF's key component that holds "sensory map", i.e., the *Place Cells*. NeRTCAM implements storage as well as inference/prediction logic consistent with the models used in [8], [16].

According to this theory, each CC can store complete models of objects in its RF, regardless of its modality or hierarchical abstraction. This contrasts traditional view in deep learning where the lower layers only learn primitive features (e.g., lines, curves) and the higher layers learn bigger and more complete features (e.g., wheel, car). These CCs communicate within and across modalities to reach consensus on the output via voting. Authors in [8]–[10] demonstrate the application of this theory through software simulations by performing environment classification and visual object recognition on MNIST handwritten digits using sequences of sensory inputs correlating with corresponding sensor locations.

Additionally, Smith has shown promising simulation results of a "macro-column" architecture [16] based on Hawkins' Reference Frame, illustrating it using the example problem of mouse navigating in a maze towards a specific target. Authors in [4] noticed the strong synergy between Hawkins' CCs and Smith's TNNs, i.e., CCs resemble TNNs with recurrence embodied in the RFs, and proposed "Cortical Columns Computing Systems" as an over-arching approach towards building neuromorphic processors. Although prior works have illustrated the potential for energy-efficient CMOS implementation of TNNs, no work has investigated the feasibility of direct hardware implementation of Reference Frames (RFs) for CCs.

Authors in [4] describe a Cortical Column (CC) (Figure 1) as consisting of two components: 1) a *Reference Frame* that maintains a "map" of the sensory information, and 2) an *Agent* that achieves goal-oriented behavior based on information from the Reference Frame and the current input signals. Agent comprises of two TNN-based compute units performing unsupervised clustering and supervised classification. Reference Frame involves three functional modules, called: *Grid Cells* (referred to as "Where" in [4]), *Place Cells* (referred to as "What" in [4]), and *Output Column*.

In the context of visual object recognition, the three modules together create models of objects by tracking locations of features on the object. *Output Column* determines the object identity after incorporating voting consensus across multiple

CCs. *Grid Cells* generate location of sensor on the object based on feedback from *Output* and the latest movement information from the agent. *Place Cells* predict object features based on the result from *Grid* and updates its model based on the actual sensory input and the feedback from *Output*. *Place Cells* is the key module responsible for learning and maintaining the "sensory map" in a Reference Frame.

As shown in Figure 1, we propose NeRTCAM as a building block for implementing this key "Place Cells" module in the RF. It not only acts as a fast storage module for storing sensory maps consisting of {feature, location, class/objectID} triplets but also supports the fundamental operations needed for prediction and inference.

*B. Ternary CAMs*

Content Addressable Memory (CAM) [17] performs fast parallel associative lookup by comparing input search query with stored content and returns the matched entries. Typically, a single matching entry is selected by implementing specific priority rules. Ternary Content Addressable Memory (TCAM) is a special type of CAM that operates on three logic states (0, 1, and "don't-care (X)") unlike traditional CAM designs using binary storage and search operations. TCAM stores entries with 'don't-care' bits and these entries are searched against binary queries. This feature enables TCAM to perform high-speed parallel search operations for partially matching content via bit-level masking, making it well-suited for applications that require fast pattern matching and routing in networking, telecommunications, database systems, etc.

One of the advantages of CAM (and thereby TCAM) is its ability to perform content-based searches in a single clock cycle. By utilizing parallel search operations across multiple memory entries simultaneously, TCAM can rapidly identify matching patterns within a large data-set. Furthermore, the inexact matching capability of TCAM due to the addition of a don't-care state allows it to model partial content matching. This makes TCAM an efficient solution for tasks such as firewall filtering and packet routing.

However, TCAM also has issues that need to be addressed in practical implementations. One challenge for TCAM is the higher power consumption [18] due to its parallel search operations and larger number of logic states. Additionally, the cost of TCAM can be high due to its specialized design and complex manufacturing process. Many works in literature have focused on efficient circuit-level implementations of TCAM [19]–[21]. In contrast, in this work, we model TCAM functionality using flip-flops (registers) and focus on designing the logic and architecture required to leverage TCAM as a building block for implementing Reference Frames. Our future work will focus on integrating circuit-optimized TCAM cells into our subsequent NeRTCAM designs.

One distinguishing attribute of proposed NeRTCAM is that instead of storing ternary entries with 'don't-care' bits and matching them with incoming binary inputs as in conventional TCAM, we store binary entries (logic states 0, 1) and provide ternary data with don't-care bits as search input. Hence, we call it "Reverse" TCAM (RTCAM). This is due to the fact that agent stores fully specified information, i.e., {feature, location, class} triplets, but during prediction and inference, it either provides one or both of location and feature with the missing entries masked by don't-care bits.

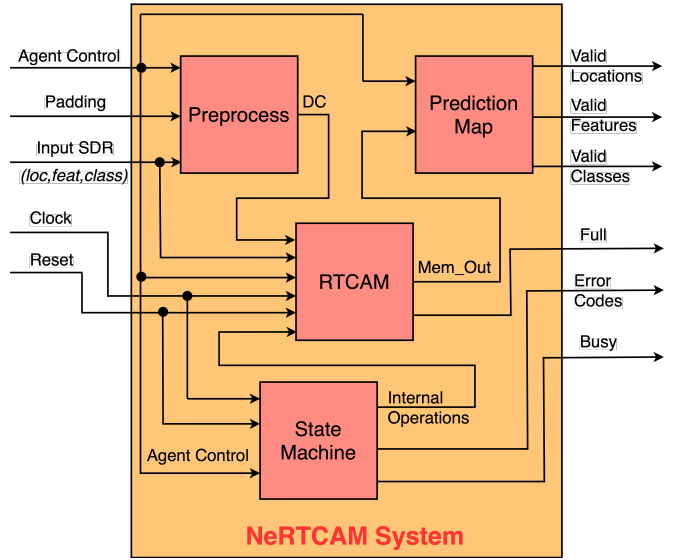

Fig. 2. NeRTCAM System consists of four main components: 1) Preprocess, 2) RTCAM, 3) State Machine, and 4) Prediction Map. Preprocess and prediction map are combinational, whereas RTCAM and state machine are sequential. The system as a whole takes in input SDR and control commands from agent and outputs valid set of feature-location-class triplets.

III. NERTCAM ARCHITECTURE

In this section, we present the proposed NeRTCAM (Neuromorphic Reverse Ternary Content Addressable Memory) architecture and describe its design and operation. First, a system overview is provided followed by description of input commands (i.e., the macro-operations queried to NeRTCAM by the agent). The system components and internal micro-operations will be discussed in detail later in Section IV.

*A. System Overview*

NeRTCAM (depicted in Figure 2) is designed to serve as a fundamental building block for implementing Reference Frame (specifically the most important "*Place Cells*" [4]), storing and retrieving information in the form of {feature, location, class} triplets. It is to be noted that it can also be used as a separate structure to which the "Place Cells" assign learned information for storage and fast retrieval during inference. NeRTCAM performs three main functions within the RF: 1) *Predict* feature and class given location of the sensor, 2) *Predict* sensor location and class given input feature, or 3) *Infer* class given location of the sensor and the sensed feature at that location. Each of these operations can output multiple matching entries. Further, NeRTCAM supports fuzzy matching as will be explained later in Section IV-A.

In order to achieve this, NeRTCAM is designed with four major components as shown in Figure 2, namely, *Preprocess*,

*RTCAM*, *State Machine*, and *Prediction Map*. NeRTCAM takes in instructions from agent in the form of control commands (*Agent Control*) and fuzziness amount (*Padding*). It also receives an input SDR (sparse distributed representation as in [8]) that is used to query matching entries based on the content stored in NeRTCAM. At the output, it generates three different vectors indicating valid set of locations, features and classes consistent with the information collected by the agent as it navigates an environment. Alongside, three status signals, namely, *full*, *error* and *busy* are also generated. High-level functionalities of the four components are described next.

The preprocess block is responsible for creating the "don't-care" (*DC*) mask pertaining to the operation the agent wishes to perform, as well as modifying it to reflect the desired fuzziness for matching. The output of this block (the DC mask) is fed to the RTCAM along with input SDR. The input SDR and DC mask are both vectors of same length with certain number of bits reserved for representing feature, location and class. Feature, location and class sections within the SDR are assumed to be 1-hot. RTCAM is the storage element possessing all the learned triplets by the agent and it outputs all matching valid outputs, i.e., class and feature/location predictions (*Mem_Out*) corresponding to the input SDR and DC mask. The state machine converts control commands from agent (macro-operations) to internal micro-operations for RT-CAM that performs memory-native functions such as lookup. Lastly, the prediction map is used to condense the multiple matching 1-hot memory output into k-hot binary vectors that represent the current set of valid features, locations, or classes to be informed to the agent.

## B. Agent Control Commands

NeRTCAM supports six control commands: CLEAR, RESET, STORE, DELETE, INFER, and PREDICT. These commands form a concise view of how an RF-based system can support an agent in the cortical columns computing paradigm. A flowchart describing the steps involved in the four operational commands (STORE, DELETE, INFER, PREDICT) is presented in Figure 3. The two simple commands (CLEAR, RESET) are omitted from the figure for brevity. All six commands are explained below.

*1) CLEAR and RESET:* CLEAR and RESET are simple housekeeping operations for NeRTCAM. The input SDR provided by agent is irrelevant for both these operations and hence is ignored by the system. When performing a CLEAR, the system clears all of its memory contents. This is typically used during initialization. It can also be used by the agent whenever it wishes to start from scratch with an empty memory (e.g., migration to a completely new environment or task that's unrelated to previously learned information).

RESET is an operation that the agent can use to force the system to 'forget' the sequential information it collected while performing an identification on a certain object. Note that an object is identified by sequentially collecting sensor location-feature information and remembering the valid set of classes (object IDs) that are consistent with all location-feature pairs observed so far. Thus, RESET is intended to be used either after a successful identification as it prepares the system to aid in the agent's observation of a different object, or after a failed identification so that the agent can either try again or move on. It can also be used whenever the agent so pleases. Note that RESET does not reset or clear the stored entries (i.e., feature-location-class triplets).

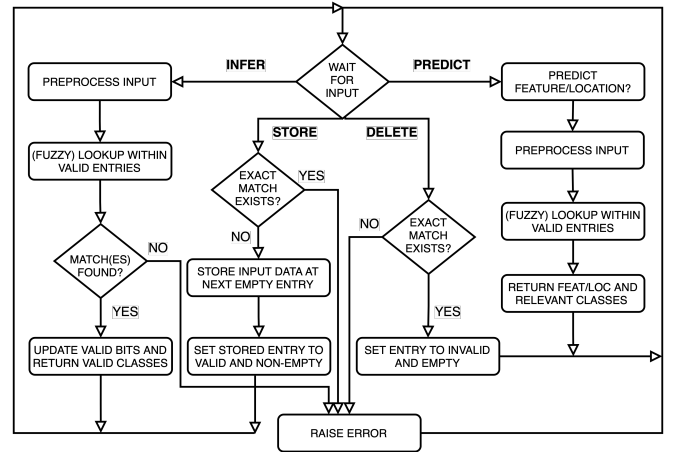

Fig. 3. Agent Control Commands Flowchart: The process of executing four key commands from the agent (STORE, DELETE, INFER, PREDICT) is illustrated in terms of the different steps involved. STORE and DELETE are performed with fully specified input SDR with feature, location and class. INFER and PREDICT are performed with partially specified input SDR along with corresponding preprocessed don't-care (DC) mask.

*2) STORE and DELETE:* STORE and DELETE are slightly more complex operations compared to RESET and CLEAR. STORE is used to add new information the agent will use for identification, into the system. When performing a store, the agent must specify exactly what feature at what location belongs to what class in the input SDR. This is done only when the agent is fully confident about its orientation within an environment. Relatively, during identification, agent's observations are more general (partial), and the system's primary goal is to match general observations with specific, remembered information. Moreover, to avoid redundancy, when performing a store, the system will check if the desired information has already been stored. If an exact match exists, the agent is notified via an error message, or else the information is stored normally. DELETE is very similar; the only difference is that it sends an error when an exact match is not present; or else that match is deleted normally.

*3) INFER and PREDICT:* These are the two main operations that assist the agent in identifying an environment or object. As part of INFER, the agent specifies a location and feature but no class. This is to replicate the act of the agent seeing a feature at a location, and then trying to remember which classes said pair belongs to. When beginning the identification process, a lookup is done on all information that is stored in the system. Any classes that contain that relevant pair are remembered as valid, and in turn, data pertaining to those valid classes will be the only data that will be used for the next lookup. During the next lookup, the classes that contain the new location-feature pair is determined out of the remaining valid classes. This continues until there is only 1 class left (successful identification) or something is seen that does not pertain to any of the relevant classes (error notification to agent).

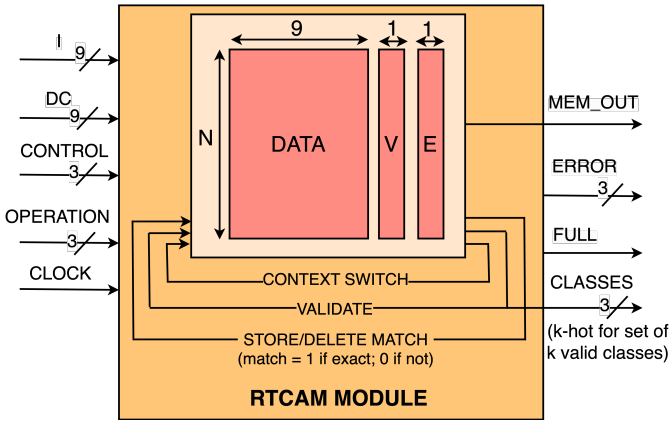

Fig. 4. RTCAM Module: This is the memory array that stores SDRs in the form of triplets - location, feature, class, along with two additional bits for valid (V) and empty (E). Assuming 'N' number of entries with 3 bits each for location, feature and class, each entry hold 11 bits in total (indicated by Nx11). It takes in input SDR (I) and agent control commands (control) from the agent, DC mask (DC) from preprocess, and internal operation (Operation) from the state machine. It generates valid location-feature-class triplets (Mem_out) and set of k valid classes as k-hot vector (classes).

The PREDICT operation has two variations that essentially perform the same desired functionality, as mentioned in Section III-A. This operation is used when the agent is unable to perform a full lookup but it has some partial information, either a location or feature, and wishes to see if said information is present in any of the valid entries. So a lookup is performed with either just feature or just location. The lookup returns which of the current valid classes contain said feature or location as well as what location or feature that input refers to (given a location, a list of features is provided and visa versa).

## IV. NeRTCAM Implementation

We now present each of the four components of the NeRT-CAM system in further detail below.

### A. Preprocess

The preprocess block is a combinational component of the system, that takes in agent control commands and padding input to generate DC vector, as shown earlier in Figure 2. For illustration simplicity, we use a 9 bit SDR (3 bits each for feature, location, and class). Hereafter, '$I$' refers to the input SDR from the agent, and '$DC$' refers to the generated 'don't-care' vector from the preprocess block.

- RESET and CLEAR: I and DC are completely ignored, and are expected to be held low.
- STORE and DELETE: $I$ = {feature, location, class} with 3-bit non-zero 1-hot vector for each; $DC$ = 9'b0. Note that for both these operations, I is completely specified and DC is ignored.
- INFER: $I$ = {feature, location, 000}; $DC$ = 000000111. Note, this assumes a padding value of 0; padding for fuzziness is explained in the next paragraph. Here, the agent is viewing a feature at a location and does not have any class data to match, so it is ignored.
- PREDICT Feature: $I$ = {000, location, 000}; $DC$ = 111000111. For this operation, the agent provides a location (with again a padding of 0). The feature and class bits are ignored as no data was provided for a match.
- PREDICT Location: $I$ = {feature, 000, 000}; $DC$ = 000111111. In similar fashion, the agent provides a feature, so the other two elements of the SDR are ignored.

Additionally, the system allows adding fuzziness to location search via padding. In order to obtain features corresponding to all nearby locations during *PREDICT Feature*, DC vector is padded with 1's around the '1' in the location part of the SDR. This is equivalent to introducing fuzziness to predict close-by features. Further, more precisely placed padding can mitigate the spatial information loss due to conversion from 2D grid to 1D vector. Note that padding in DC is only done for location part of the SDR as it does not make sense to introduce fuzziness to feature or class.

As an example, suppose the location part of the SDR is now 5 bits long and is equal to 00100, and the padding amount is 1. In this case, when performing a lookup, we need to search location values 00100, 01000, and 00010 to register as a match. This is achieved by changing the resulting DC vector's location section from all 0s (as in *PREDICT Feature* bullet point above) to 01110.

### B. RTCAM

Figure 4 illustrates the RTCAM module with its inputs and outputs. It consists of four sub-modules as described below:

*1) Memory Module:* This is currently implemented using flip-flops (registers). Each entry in the memory module consists of (f+l+c+2) bits where 'f', 'l', and 'c' refer to the number of bits for feature, location and class respectively. The additional 2 bits denote valid (valid class) and empty (empty entry). Note that the corresponding input SDR and DC vector will have (f+l+c) bits. As shown in Figure 4, it has two internal outputs, *mem_out* and *valid_entry*. The latter is used by the state machine to make decisions, and the former is all the data sent to the prediction map and must be condensed during a prediction. The other output is *infer_class_out* which is a k-hot vector of the valid classes after a successful inference. The valid bits are automatically updated each cycle with a lookup.

*2) Validation Module:* Validation is crucial to this system as it keeps track of the valid classes throughout multiple identification cycles in sequence. This step is performed after each successful lookup portion of INFER which only marks entries valid with matching feature-location pair. But, the next search will not have that same location and feature, so all classes from the result of a previous search must have ALL their entries marked as valid. To do this, we perform an internal validation operation.

For this operation, after a lookup, we create a bit representation of all valid classes, 1 bit high for each class's place in the 1-hot bit structure. We then pad it with 0s until it's the same size as an input for I and pad it with 1s for DC. We then reset the validation bits and perform another 'lookup'. The specified input is used to perform a search that ignores all bits except those that would be high to represent the valid classes. As classes are stored as 1-hot, with all 0s not valid, this means entries with classes having a 1 bit in any of the marked bits, will be marked as valid. Between each successful INFER, we output the condensed class vector so that the agent knows that classes are currently valid after this operation.

*3) Context Switch Module:* The context switch feedback loop is used to denote the system level functionality of being able to determine when the agent has started viewing a new object without performing a reset. In this case, a normal lookup fails, and rather than saying the entire inference fails, we reset the valid bits, try again, and see if there are any valid entries. If yes, then we know that the agent has switched to viewing a new object and can validate accordingly. If not, then we know that the inference simply failed, as the agent tried to perform an inference with unlearned data.

*4) Store/Delete Match Module:* When storing a new value, we need to ensure the data requested to be stored is not already in the RTCAM. For a store, we first perform a lookup of the store itself, and if we get any exact matches (1s in the valid bits of the RTCAM), then we simply reset the valid bits and await the next input. If we get no exact matches, then we find the next empty entry in the RTCAM and store the data there, setting its empty bit to 0 (now indicating not empty). We then reset the valid bits and wait for the next input. If there are no empty entries in the RTCAM, then we fail the store and notify the agent. Delete operates similarly, except finding an exact match results in the desired operation (i.e., delete) and no match results in failure of the operation and resultant error notification to the agent.

## C. State Machine

The agent operations vary widely in terms of complexity. To maximize clock frequency, some of them are performed across multiple clock cycles. Some agent operations may require multiple clock cycles involving a sequence of internal "micro-operations". A set of internal operations ("micro-ops") is developed to aid the implementation of agent operations ("macro-ops") that get executed via a sequence of single-cycle internal operations. These internal operations (micro-ops) are generated by a state machine (Figure 5) and are listed below:

- clear: clears all the information stored in the RTCAM and sets all the empty bits high.
- reset: resets all valid bits back to 1, in turn making the system 'forget' what the agent has observed in this identification cycle/the previous lookup.
- store: adds an entry to the RTCAM and sets its empty bit low and valid bit high.

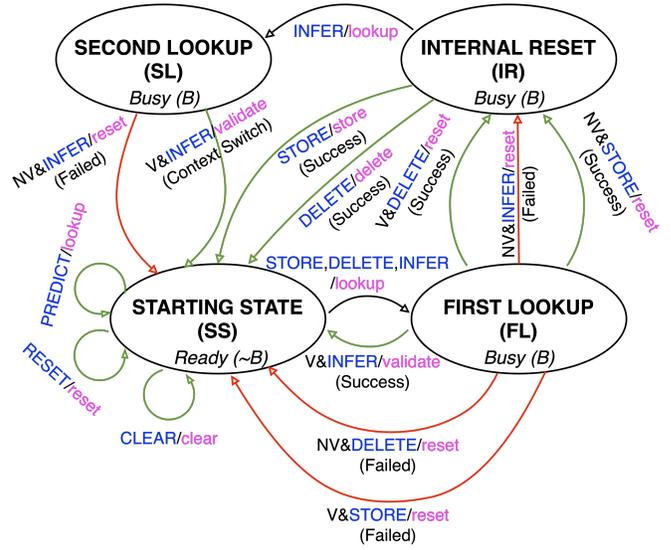

- delete: removes an entry from the RTCAM and sets the empty bit high.
- lookup: performs a query into the RTCAM; valid bits are updated and stored accordingly.
- validate: given a previously successful lookup, mark all entries that contain the valid classes as valid and output said classes.

Fig. 5. State machine: It consists of four states (SS, FL, IR, and SL). Each transition shows agent control command in blue followed by internal operation in pink, along with valid (V) and not valid (NV) bits from memory output. Status of the control command such as Success, Fail or Context Switch is provided in parentheses. Green lines indicate success and red lines indicate failure of the command.

The state machine is also responsible for error generation. It generates four types of errors to be sent to the agent: 1) *Delete_Failed* - no matching entry, 2) *Store_Failed*: data already present, 3) *Infer_Failed* - the system cannot continue as a specified lookup returned with no valid entries remaining in the RTCAM, and 4) *Context_Switch* - the system realizes that the agent has moved onto looking at a new, learned object and has adjusted accordingly.

Figure 5 shows the state transition diagram. The state machine consists of four states: *Starting State (SS)*, *First Lookup (FL)*, *Internal Reset (IR)*, and *Second Lookup (SL)*. Each transition is marked by the input agent control command (blue capitalized), followed by the generated internal operation (pink) to be executed by the RTCAM memory. RTCAM, in turn, provides valid (V) or not valid (NV) signals for its memory output, that the state machine uses to determine its transitions. Green lines indicate successful control commands and red lines indicate failure of those commands. SS is the default state and the only state in which the NeRTCAM accepts new input from an agent; system is considered busy (B) in other states. CLEAR, RESET, and PREDICT commands execute in one clock cycle as denoted by the self-loops in *SS*. STORE and DELETE require a lookup followed by a reset of the valid bits, taking two cycles in the event of failure, and an additional cycle for the actual store/delete operation if successful. INFER is executed through a lookup followed by validate, thus finishing in two cycles if successful. On failure, it performs a reset instead of validate after lookup, and takes two additional cycles - one cycle for a second lookup and another cycle for returning with failure or success (resulting in context switch as discussed in IV-B3). Figure 5 clearly depicts these transitions for every case. An example is provided below for ease of interpretation of the state transition diagram.

Beginning in the *SS* state, consider agent provides an INFER command. The system transitions to *FL* with busy (B) asserted and executes a lookup internal operation in the RTCAM to find matching entries. On success, the system transitions back to *SS* and executes a validate operation in the RTCAM to generate the output. On failure to find any valid entries (indicated by NV), the system checks for a context switch to a new object. This is done by first transitioning to *IR* and performing a reset of the valid signals, followed by a second lookup in the next cycle in the *SL* state. The reset enables the system to detect new objects. If valid outputs are found after second lookup, a successful context switch is performed by executing validate in the RTCAM and returning back to *SS*. On failure, it executes reset instead of validate, returns to *SS* and notifies an "Infer_Failed" error to the agent. Note that the busy signal is continuously asserted until the system reaches *SS*, when it is finally de-asserted. Overall, on a context switch, the operation takes four clock cycles owing to four state transitions.

### D. Prediction Map

The prediction map (as illustrated earlier in Figure 2) is another combinational block of the system. Its primary purpose is to output the desired information specified by the control. It condenses the data given by the *mem_out* signal in the same way that the validate system condenses valid bits. However, it has the capability to do this for all 3 sections of the SDRs. The idea is to predict the set of valid features/locations/classes, with a one-hot class output indicating a strong prediction for one particular class that the agent can choose to use (i.e. a classification). Otherwise, it just indicates the set of valid classes remaining at this point in the prediction. The same form of compaction is needed to select only the valid entries from the previous stage and condense it into a singular k-hot vector, that could then be used by the agent to distinguish which features/locations are expected to be seen next.

If the agent operation specified is not a type of PREDICT, then nothing is given as output. Depending on the type of predict, the location output or feature output will be held low. For PREDICT_FEATURE, the location output is low and vice versa. This is because after performing the lookup, all valid entries in the RTCAM are going to have the same location (for a predict feature), which is the location provided in the input. So, there is no reason to output this. Also, in case the generated output is 0 (not purposefully held 0), it implies the prediction failed and the agent is trying to predict a location/feature pair that is not valid in the given identification cycle.

TABLE I
7NM CMOS POST-SYNTHESIS PPA RESULTS FOR NERTCAM WITH THE NUMBER OF ENTRIES SCALED FROM 64 TO 1024. EACH ENTRY STORES 165 BITS = 163-BIT SDR (128 BITS FOR FEATURE, 25 BITS FOR LOCATION, 10 BITS FOR CLASS) + 1 VALID BIT + 1 EMPTY BIT (BASED ON MNIST BENCHMARK REQUIREMENTS FROM [8]).

| Number of Entries | Power (mW) | Latency ($\mu$s) | Area ($\mu m^2$) |
|---|---|---|---|
| 64   | 26.30  | 2.06 | 9,468.73    |
| 128  | 53.02  | 2.26 | 18,912.53   |
| 256  | 106.54 | 3.08 | 38,961.76   |
| 512  | 213.39 | 6.91 | 77,441.56   |
| 1024 | 400.23 | 9.18 | 153,914.94  |

## V. RESULTS & EVALUATION

### A. Experimental Setup & Methodology

Parameterized RTL designs for NeRTCAM and all its components are implemented in SystemVerilog. Functional RTL simulation is performed using Cadence Xcelium and logic synthesis using Cadence Genus with open-source 7nm predictive ASAP7 PDK [22]. Clock frequency used is 100 kHz to achieve real time operation similar to biological scales.

For correctness testing, rigorous testbenches are created to test every component including every state transition for the state machine. Post-synthesis power measurement is done using switching activity file generated through functional RTL simulation. The test vectors provided for power measurement are sampled from MNIST SDRs as used in [8]. Authors in [8] generate a 128-bit feature vector on a 5x5 grid using a CNN, for each MNIST digit. Aligning with this approach, we use 25 bits for location, 128 bits for feature, and 10 bits for class (for 10 handwritten MNIST digits). This implies a total SDR length of 163 bits. This SDR configuration is used to generate the post-synthesis results (not 3 bits each, which was used as an illustrative example in the earlier sections). Lastly, note that the feature vectors in [8] are not 1-hot but are 19-hot, however our system can be extended to support k-hot feature vectors as long as locations and classes are still 1-hot.

### B. 7nm PPA Evaluation

NeRTCAM is able to successfully demonstrate sequential inference on MNIST as presented in [8]. The SDR bit lengths for post-synthesis evaluation are determined using the same MNIST example from [8] as mentioned earlier. Accounting for the two additional valid and empty bits, each entry consists of 165 bits in total. The number of entries is varied from 64 to 1024, totaling five design configurations.

Table I presents the post-synthesis 7nm power, performance (latency) and area results for varying storage sizes (number of entries) of NeRTCAM. Area and power scale almost linearly with respect to the number of entries, almost doubling with every 2x increase in size, whereas critical path latency scales much more gradually. The smallest assessed NeRTCAM with 64 entries incurs less than 0.01 mm$^2$ area and 26.3 mW power. Even NeRTCAM storing 1024 entries consumes just 0.15 mm$^2$ area, 400.23 mW power, and 9.18 $\mu$s latency which is very amenable for real-time operation. This area footprint is about 0.1% of typical mobile SoCs where memory dominates die area, thus demonstrating the potential and feasibility of implementing multiple NeRTCAM modules in parallel or in cascaded fashion to achieve highly scalable compute fabric.

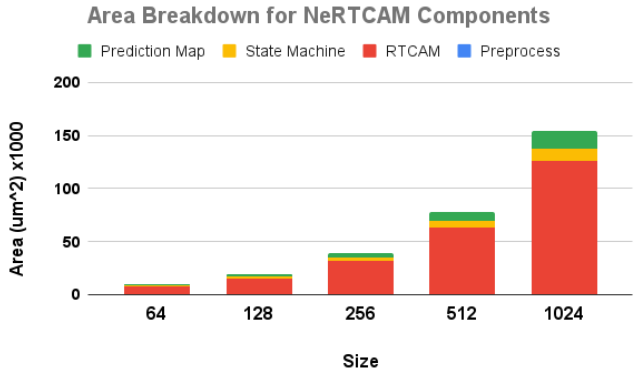

Fig. 6. Hardware Complexity (Area) Breakdown for NeRTCAM in terms of its four components: Preprocess, RTCAM, State Machine, and Prediction Map. RTCAM consumes 82.2% of the area, followed by 10.5% for Prediction Map and 7.3% for State Machine. Preprocess incurs negligible complexity.

Authors in [8] illustrate a few key desirable features for their proposed neural network (GridCellNet) inspired from Grid Cells and Reference Frame: 1) *Learning with Arbitrary Sequences*: GridCellNet significantly outperforms similar LSTM and KNN implementations when MNIST inference is performed using arbitrary sequences of feature-location inputs. In fact, in this challenging biologically motivated scenario, KNN is unable to learn well and reaches only ∼30% accuracy, whereas LSTM and GridCellNet achieve ∼65% and ∼80% respectively. 2) *Few-shot learning*: With arbitrarily sequenced inputs, GridCellNet achieves above 70% accuracy with only 5 training samples per class and reaches 80% with 20 samples per class, while consistently outperforming LSTM and KNN implementations. 3) *Rapid inference with partial sequences*: GridCellNet can successfully classify majority of the MNIST digits in under 10 sensations, i.e., it doesn't require all 25 sensations (at 25 locations) to infer the digit. These results from [8] indicate that NeRTCAM with 500 entries (10 sensations per sample, 5 samples per class, and 10 classes) can successfully achieve >70% accuracy for arbitrary sequences of inputs outperforming LSTM and KNN implementations, while consuming just above 200 mW power and 0.08 mm$^2$ area. Note that GridCellNet's largest configuration (25 sensations per sample, 20 samples per class, and 10 classes) results in 5000 entries for NeRTCAM, which is an overspecification and unnecessary for successful MNIST inference.

Lastly, Figure 6 illustrates the hardware complexity breakdown in terms of area, for the four major components of NeRTCAM. RTCAM memory consumes majority (82.2%) of the area as expected, followed by 10.5% for Prediction Map and 7.3% for State Machine. These three components scale almost linearly with the number of entries. In contrast, Preprocess barely scales with number of entries and incurs negligible complexity. The dominant contribution to hardware complexity from RTCAM memory module implies replacement of flip-flops with custom TCAM cells in RTCAM can potentially provide significant improvements in PPA.

## VI. Conclusions & Future Work

Cortical columns with Reference Frames are the fundamental units of intelligence within the brain. Prior works have shown the application potential for Reference Frame-based models in performing movement-based visual object recognition and environment classification through software simulation. This work serves as the first attempt towards CMOS implementation of such a Reference Frame and proposes a Content Addressable Memory (CAM)-based building block - Neuromorphic Reverse Ternary CAM (NeRTCAM). NeRTCAM consists of four major components, Preprocess, RTCAM, State Machine and Prediction Map and can perform sequential prediction and inference as the sensor moves in an environment. Out of these, RTCAM incurs majority of the hardware complexity. NeRTCAM system storing 1024 entries (about 100 entries for each digit) can perform successful inference through sequential samples of MNIST [8] while incurring only 0.15 mm$^2$ area, 400 mW power and 9.18 $\mu$s latency. Such sequential inference, especially with arbitrary sequences of sensations from an image as typically occurs in biology, is challenging for traditional ML algorithms like LSTMs and KNNs. The main goal of this work is to demonstrate the feasibility of direct CMOS implementation of a biologically plausible CAM-based Reference Frame and thereby serve as the initial baseline for future follow-on works on RF-based neuromorphic processors.

In terms of future work, an immediate next step is to integrate specialized circuit-level TCAM cells into RTCAM, replacing the current highly expensive and power-hungry flip-flops. As noted earlier, this modification can result in significant improvement in area and power consumption of NeRTCAM. A vast body of prior literature on TCAM circuit implementations can be leveraged for this. Another issue is the 1-hot assumption of SDRs which exhibits inefficient scaling. While we believe many current applications can be converted to 1-hot, efficient mechanisms to handle k-hot SDRs (especially locations and classes) can be devised for improved scalability. Additionally, NeRTCAM only serves as a building block for the "sensory map" storage within Reference Frame. Other components such as "Grid Cells" and the overall Reference Frame architecture including the learning algorithm need to be developed to build a complete system.

## Acknowledgment

This work benefited from collaboration and discussions with Dr. Niels Leadholm and Dr. Viviane Clay from Numenta, following their guest lecture in Neuromorphic Computer Architecture course (18-743) at Carnegie Mellon University.